\documentclass[a4paper,11pt]{article}
\pdfoutput=1 

\usepackage{jinstpub} 

\usepackage{lineno}
\usepackage{graphicx}
\usepackage{amsmath}
\usepackage{siunitx}
\usepackage{hyperref}
\usepackage{comment}
\usepackage{float}

\usepackage{booktabs}
\usepackage{subcaption}
\usepackage{cleveref}

\title{\boldmath High Density Hybridisation Using ENIG Bumping and Anisotropic Conductive Films}

\author[a]{Haripriya Bangaru,}
\author[b]{Giovanni Calderini,}
\author[a]{Dominik Dannheim,}
\author[a]{Rui De Oliveira,}
\author[a,1]{Ahmet Lale,\note{Corresponding author}}
\author[d]{Moritz Lauser,}
\author[c]{Mateus Vicente Barreto Pinto,}
\author[a,2]{Peter Svihra\note{Now at Faculty of Nuclear Sciences and Physical Engineering, Czech Technical University.},}

\affiliation[a]{CERN,\\Geneva, Switzerland}
\affiliation[b]{LPNHE-Paris, Centre National de la Recherche Scientifique,\\Paris, France}
\affiliation[c]{Université de Genève,\\Geneva, Switzerland}
\affiliation[d]{KIT - Karlsruhe Institute of Technology,\\Karlsruhe, Germany}

\emailAdd{ahmet.lale@cern.ch}

\abstract{Fine-pitch hybridisation processes are essential for next-generation pixel detectors and high-density microelectronic assemblies. Conventional bump-bonding techniques, although reliable, remain costly and difficult to implement for single-die applications.

In this work, we present flip-chip hybridisation results combining Electroless Nickel Immersion Gold (ENIG) bumping with Anisotropic Conductive Film (ACF) bonding, both developed in-house. This method enables fine-pitch interconnections without requiring wafer-level processing.

The feasibility of the ENIG–ACF process for functional devices was demonstrated by hybridising functional Timepix3 ASIC and Ti-LGAD (Trench Isolated Low-Gain Avalanche Detector) sensors with 55~µm pitch. The approach was then extended to test structures with 25~µm pitch,  using aligned-particle ACFs. The analysis revealed that the main limitation originated from the mismatch between ACF thickness and bump height, underlining the importance of interface geometry optimisation. To address this, a dedicated simulation program was developed to model the isotropic growth of ENIG bumps and to determine the optimum bump height as a function of ACF thickness and pixel-matrix geometry.

The results provide valuable guidelines for future high connection density assemblies and demonstrate the potential of the ENIG–ACF process as a scalable, low-cost alternative to conventional bump-bonding techniques.}


\arxivnumber{} 

\notoc

\begin{document}

\maketitle

\thispagestyle{empty}
\setcounter{page}{0}

\section{Introduction}

Conventional bump-bonding techniques, although well established, rely on wafer-level processes that are expensive and challenging to implement for single-die hybridisation. This limitation is particularly important in research and development contexts, where chips are often produced in small quantities, on multi-project wafers (MPWs). To overcome these constraints, alternative interconnection methods based on anisotropic conductive materials, and compatible with single-die processing are being explored as part of the CERN EP R\&D program on technologies for future experiments \cite{EPreport} and within the AIDAinnova \cite{AIDA} and DRD3 \cite{DRD3} collaborations.

Within this framework, an innovative hybridisation approach combining Electroless Nickel Immersion Gold (ENIG) bumping with Anisotropic Conductive Film (ACF) interconnection is under development \cite{Ahmet}\cite{Hari}\cite{Alex}. This technique aims to provide a flexible, cost-effective, and scalable solution for fine-pitch flip-chip assemblies. The ENIG process, performed in-house, enables the formation of metallic bumps on individual dies without the need for complex lithographic processes. The ACF bonding uses conductive particles embedded in a thin adhesive layer to establish electrical connections. These two techniques, when applied together, provide a promising cost-effective method for fine-pitch hybridisation.

This study focuses on the experimental evaluation of this ENIG-ACF hybridisation process. The first part presents the flip-chip assembly of a functional Timepix3 ASIC and two Ti-LGAD (Trench Isolated Low-Gain Avalanche Detector) sensors at 55~µm pitch, demonstrating the feasibility of the method on functional devices. The second part investigates the scalability of the process down to 25~µm pitch using dedicated test structures, assessing the impact of ACF thickness, bump geometry, and particle alignment on connection yield. Finally, a newly developed simulation program is introduced to model ENIG-bump growth and optimise bump height with respect to ACF thickness and chip geometry.

\section{Fine-Pitch Hybridisation of Functional Chips Using ENIG Bumping and ACF}

\begin{figure}[htbp]
\centering
\begin{subfigure}{0.32\textwidth}
    \centering
    \includegraphics[width=\textwidth]{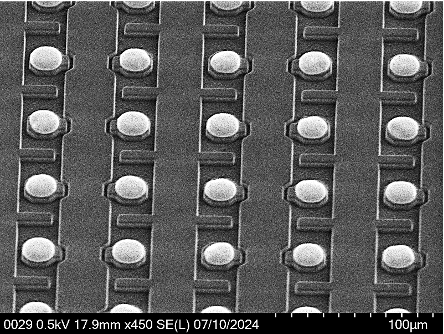}
    \caption{}
\end{subfigure}
\begin{subfigure}{0.29\textwidth}
    \centering
    \includegraphics[width=\textwidth]{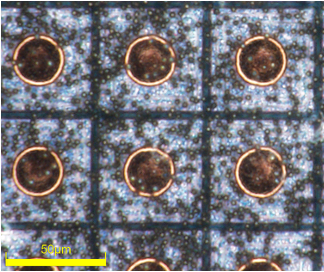}
    \caption{}
\end{subfigure}
\hfill
\begin{subfigure}{0.37\textwidth}
    \centering
    \includegraphics[width=\textwidth]{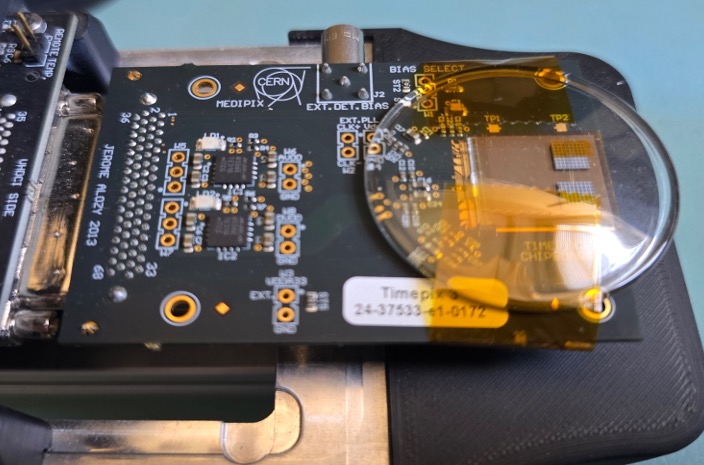}
    \caption{}
\end{subfigure}
\hfill
\caption{Timepix3/Ti-LGAD hybridisation process and module pictures. a) ENIG bumps on Timepix3 chip (Scanning Electron Microscope (SEM) picture), b) Ti-LGAD sensor after ACF lamination (optical microscope), c) Timepix3/Ti-LGAD hybrid wire-bonded to PCB.}
\label{fig:1}
\end{figure}

To evaluate the performance of the ENIG-ACF hybridisation process previously presented \cite{Ahmet}\cite{Hari}, tests were done on functional chips. A Timepix3 ASIC with a 55~µm pitch and a 256 × 256 pixel matrix was hybridised with two 64~×~64 pixels Ti-LGAD sensors (FBK–AIDAinnova production), also featuring a 55~µm pitch.
The Ti-LGAD sensors already had a 5~µm-thick, 25~µm-diameter UBM (Under Bump Metallization) layer previously deposited on their pads. The ASIC, on the other hand, did not have any UBM on its pads. Therefore, bumps were created on them using the ENIG process (Fig. \ref{fig:1}a). The bumps obtained were 6~µm high above the pad, with a diameter of 19~µm.

For the flip-chip hybridisation, an ACF of 18~µm thickness containing 3~µm-diameter nickel-coated conductive particles (CP) with a density of 71,200~CP/mm² was used. The ACF was laminated on the Ti-LGAD sensors at 90~°C under a pressure of 3.2~kgf (Fig. \ref{fig:1}b), followed by bonding at 160~°C under 20~kgf for 2~s using a SET ACCµRA100 flip-chip bonder \cite{SET}. Two sensors were connected side-by-side to the same ASIC.
The hybrid detector obtained was then attached to a dedicated readout PCB, and electrical connections between the hybrid and the PCB were established by wire-bonding (Fig. \ref{fig:1}c).

First, an electrical leakage-current measurement as function of applied bias voltage was performed to verify the proper operation of the sensors. One of the two sensors showed a short circuit, while the other displayed the expected diode-like I(V) behaviour. As the sensors had not been electrically tested prior to hybridisation, it was not possible to determine whether the failure occurred before or during the process. Therefore, subsequent characterisation focused on the functional sensor.

\begin{figure}[H]
\centering
\begin{subfigure}{0.28\textwidth}
    \centering
    \includegraphics[width=\textwidth]{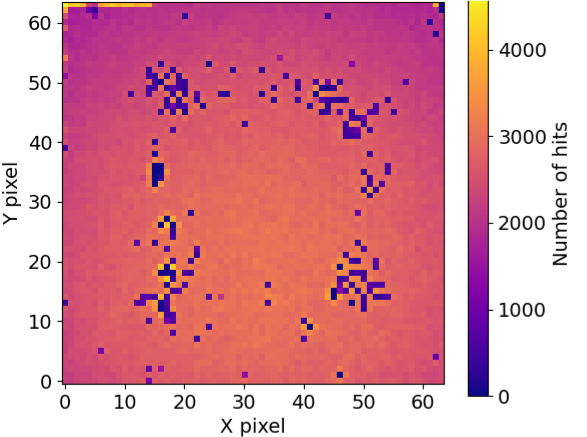}
    \caption{}
\end{subfigure}
\hfill
\begin{subfigure}{0.36\textwidth}
    \centering
    \includegraphics[width=\textwidth]{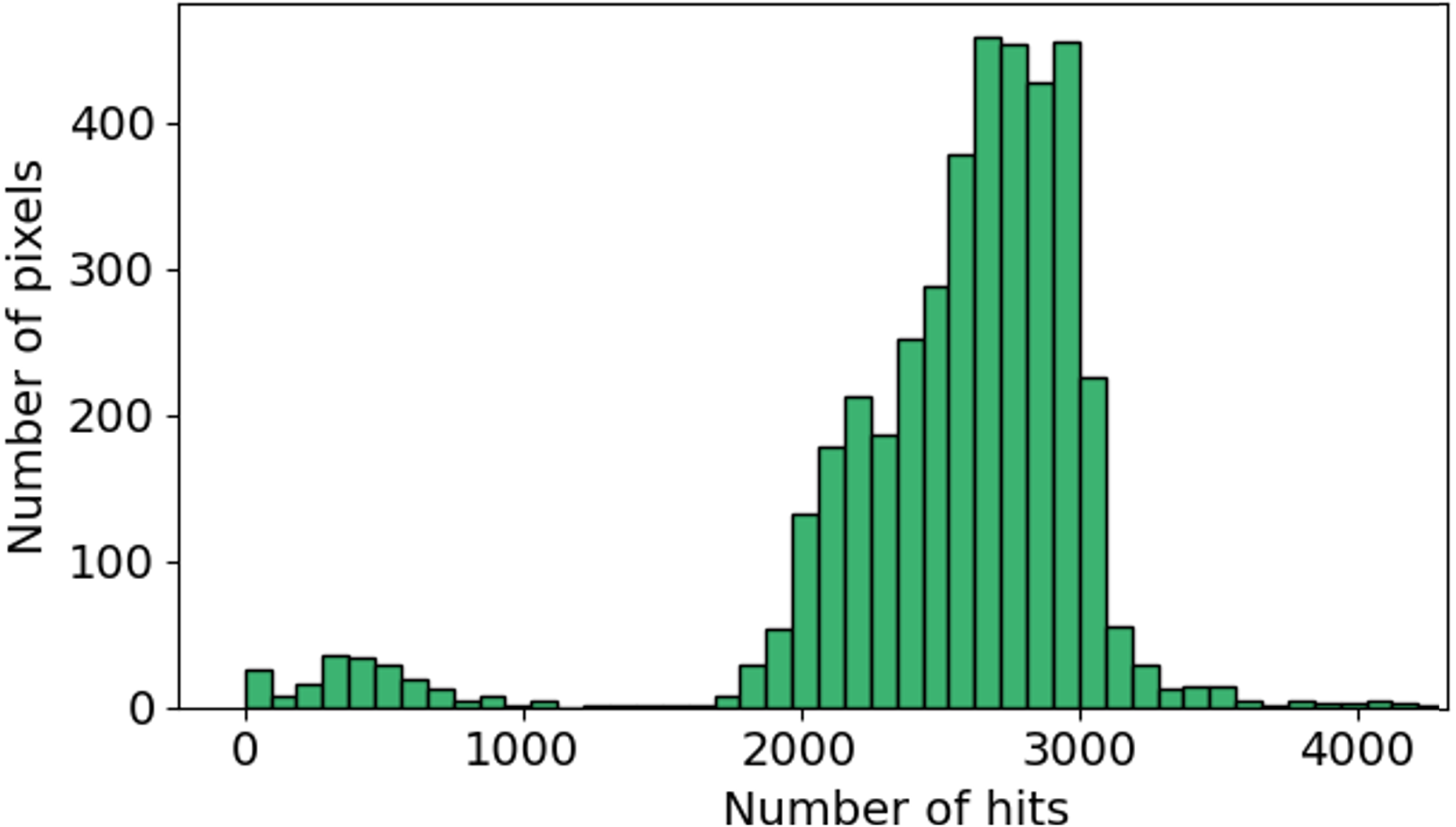}
    \caption{}
\end{subfigure}
\hfill
\begin{subfigure}{0.2\textwidth}
    \centering
    \includegraphics[width=\textwidth]{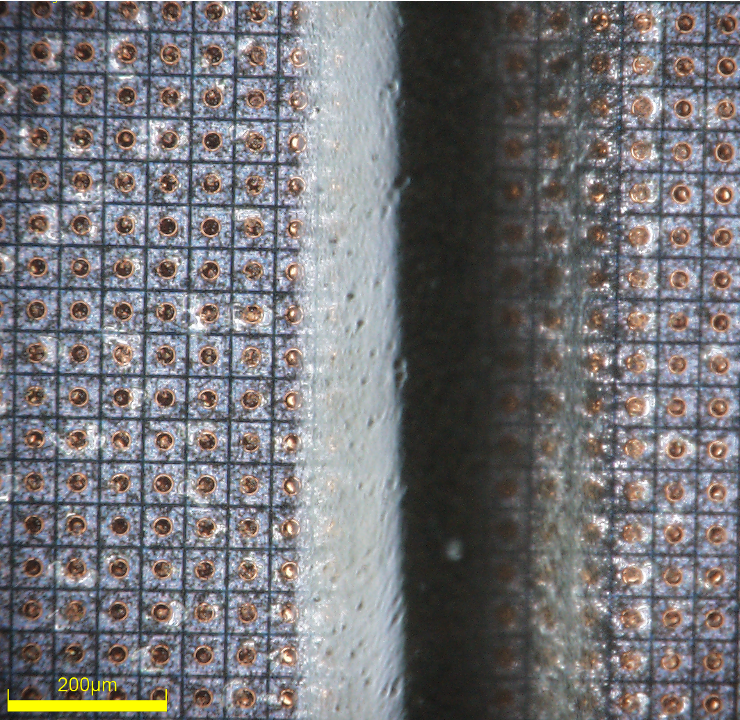}
    \caption{}
\end{subfigure}
\hfill
\caption{Timepix3/Ti-LGAD hybrid characterisation pictures. a) Hitmap obtained with a strontium-90 source, b) Histogram representing the number of hits per pixel, c) optical-microscope picture of the lamination issue in the vacuum-groove region.}
\label{fig:2}
\end{figure}

To evaluate the pixel connection yield, the hybrid was exposed to a strontium-90 radioactive beta source under a reverse bias of 100~V applied to the backside of the Ti-LGAD. The measured leakage current was about 100~µA during the test. The resulting hit map (Fig. \ref{fig:2}a) revealed the pixel response. To further analyse these data, a histogram of the number of recorded hits per pixel was plotted (Fig. \ref{fig:2}b). Two distinct peaks were observed: the first one at lower hit counts corresponds to pixels with a weaker response (considered not fully connected or defective), and the second one to pixels with a strong response, interpreted as well-connected.

Based on these data, the overall connection yield was estimated to be 95~\% of the 4096 pixels (195 weakly responding pixels). Most of the weakly responding pixels formed a 2~mm~×~2~mm square at the centre of the pixel matrix (174 pixels concerned). We attribute this to the vacuum grooves on the surface of the bonding tool used for both ACF lamination and flip-chip bonding. These grooves caused a local accumulation of ACF material during lamination, leading to poor connections for pixels located in those regions (Fig. \ref{fig:1}c).
Excluding the pixels affected by this lamination issue, 99.5~\% of the remaining  pixels were well connected (21 pixels outside the groove-regions are weakly responding). Future tests using a bonding tool without grooves are therefore expected to achieve connection yields close to 99~\% under the same process conditions.

In conclusion, this hybridisation test shows the feasibility of hybridising 3.5~×~3.5~mm² pixel matrices with a 55~µm pitch using ACF technology combined with ENIG bumping. However, the lamination step should be improved by using an adapted tool.

\section{25~µm Pitch Hybridisation Using ACF and ENIG Bumping}

\begin{figure}[htbp]
\centering
\begin{subfigure}{0.28\textwidth}
    \centering
    \includegraphics[width=\textwidth]{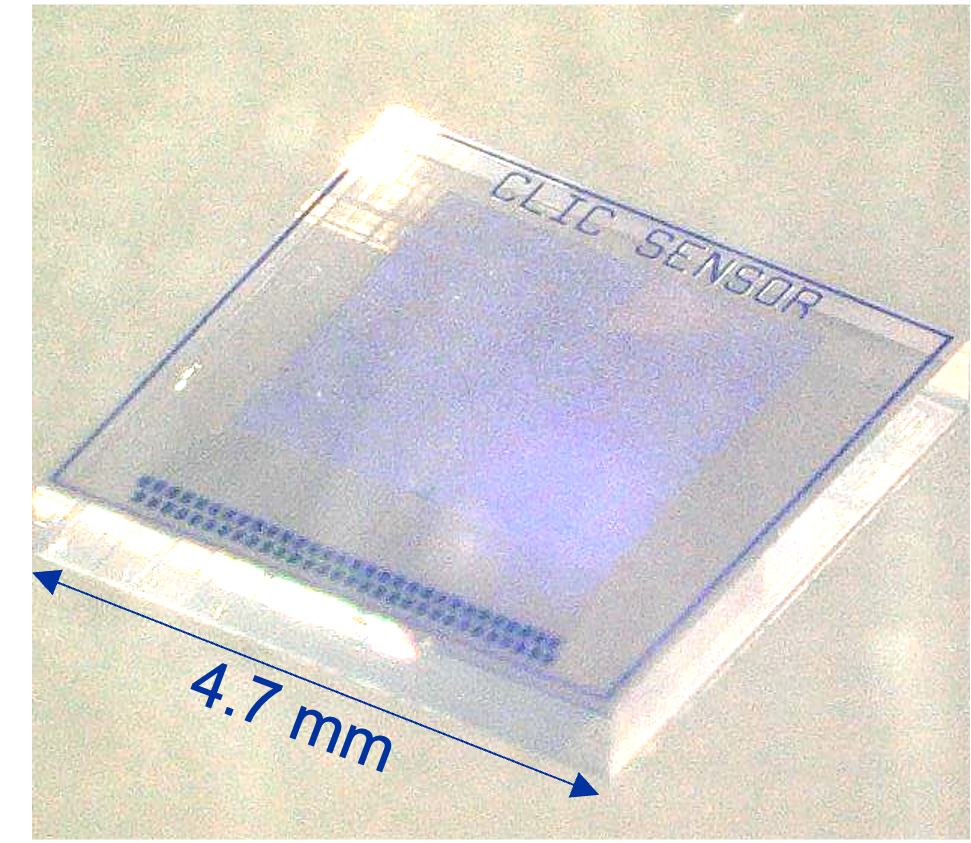}
    \caption{}
\end{subfigure}
\hfill
\begin{subfigure}{0.21\textwidth}
    \centering
    \includegraphics[width=\textwidth]{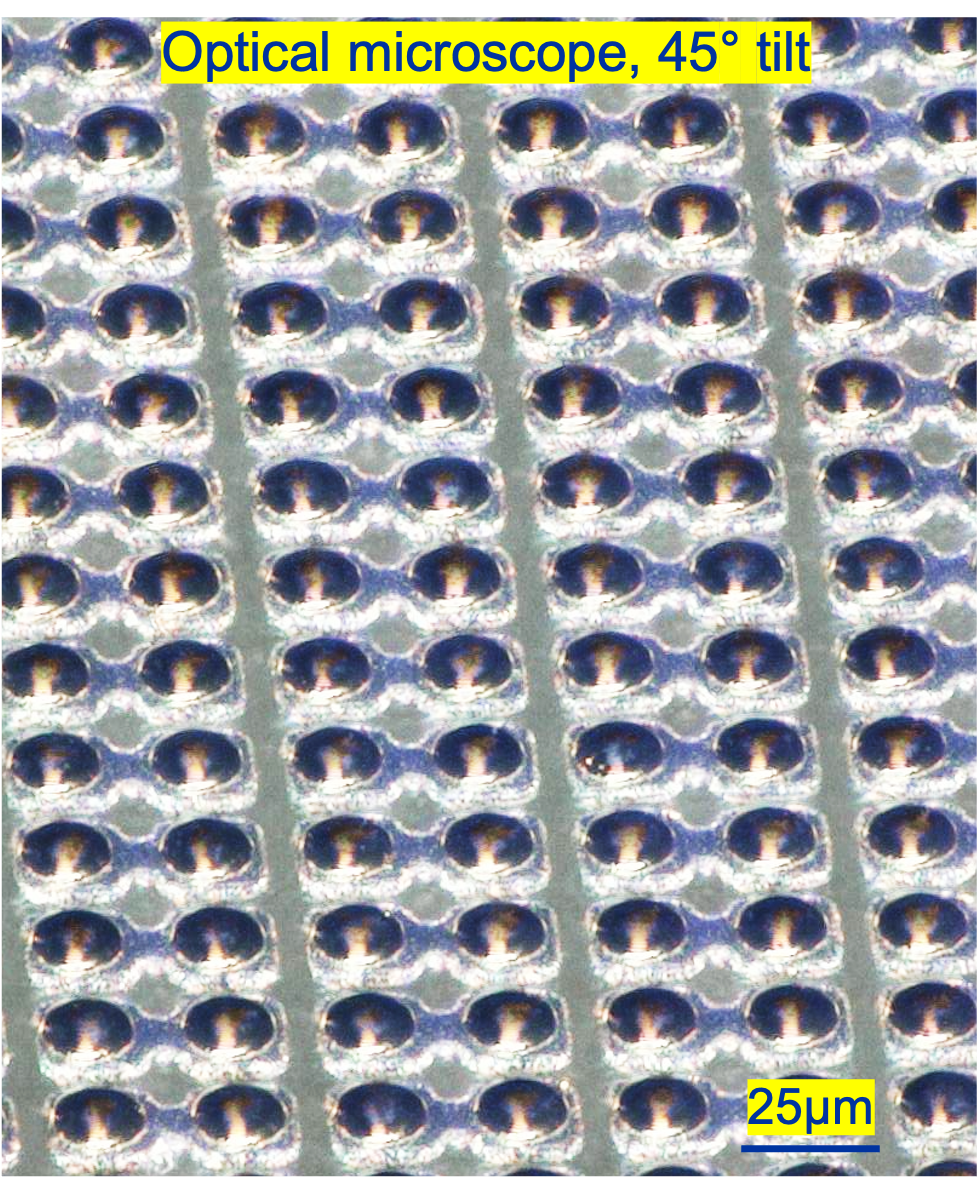}
    \caption{}
\end{subfigure}
\hfill
\begin{subfigure}{0.25\textwidth}
    \centering
    \includegraphics[width=\textwidth]{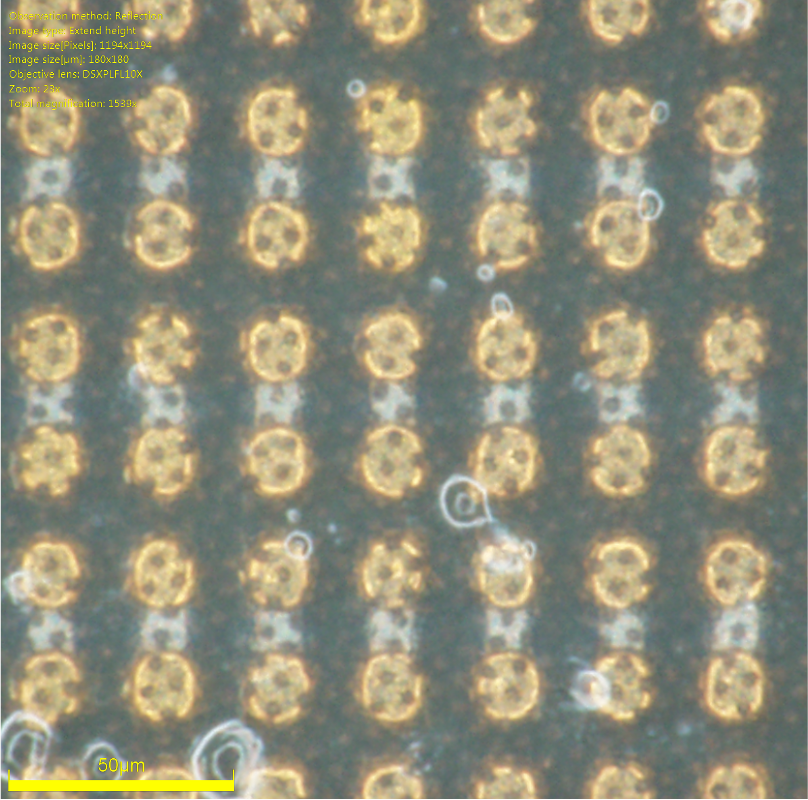}
    \caption{}
\end{subfigure}
\hfill
\caption{(Optical-microscope pictures of test-chip hybridisation process. a) Quartz test chip, b) ENIG bumps realised on the test chips, c) Test chip after ACF lamination.}
\label{fig:3}
\end{figure}

To assess the compatibility of the ENIG and ACF-based hybridisation process with chips featuring pitches below 55~µm, experiments were carried out on custom-designed interconnect-test chips with 25~µm pitch \cite{Calderini}. The chip dimensions are 5 mm~×~4.7 mm with a matrix size of 128~×~128 aluminium pads (Fig. \ref{fig:3}a). These test structures are made of quartz (transparent), allowing visual confirmation of alignment and evaluation of ACF particle density. Daisy-chain structures are implemented on them, allowing evaluation of the interconnection yield by measuring the resistance of the electrical chains after flip-chip bonding. 

First, ENIG bumps were realised on the chips using the in-house ENIG process. The bump height obtained was 4.5~µm~±~0.2~µm on both chips (Fig. \ref{fig:3}b).

The flip-chip process was then performed as follows: first, an anisotropic conductive film was laminated at 80~°C under a force of 1~kgf on one of the two chips (Fig. \ref{fig:3}c). The ACF used has a thickness of 18~µm and contains nickel-coated conductive particles with a diameter of 3.2~µm and a density of 28,000~CP/mm². Unlike the ACFs typically used in our previous tests, this material featured aligned conductive particles. The pitch of the particles position is about 6.4~µm in a triangular matrix shape. The main advantage of the aligned particles in the ACF is that it increases the connection probability in small-pitch and small-pad configurations, compared to ACFs with randomly distributed particles, which often exhibit particle clustering and void formation (areas without particles).

\begin{figure}[htbp]
\centering
\begin{subfigure}{0.385\textwidth}
    \centering
    \includegraphics[width=1\textwidth]{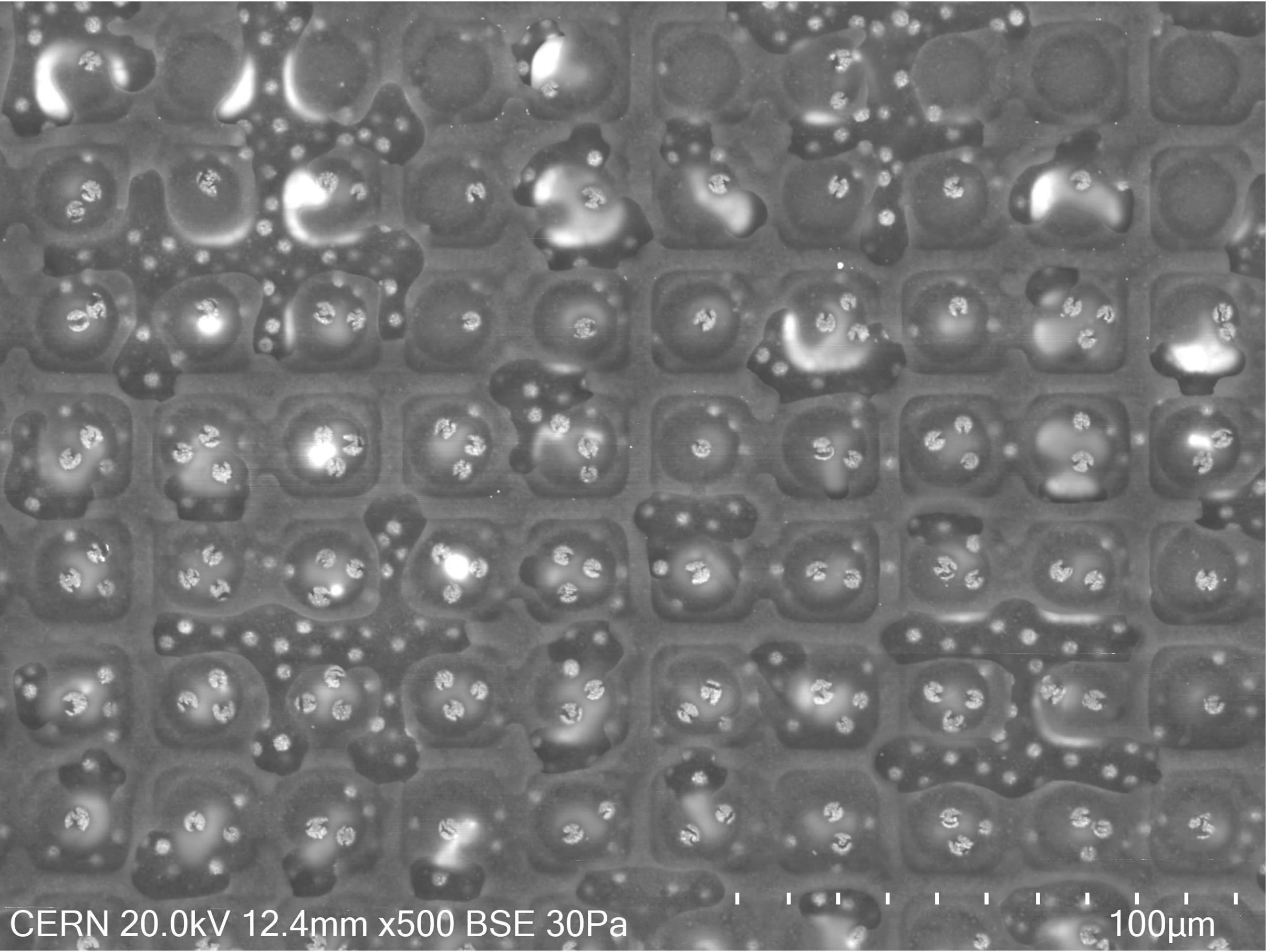}
    \caption{}
\end{subfigure}
\hfill
\begin{subfigure}{0.288\textwidth}
    \centering
    \includegraphics[width=\textwidth]{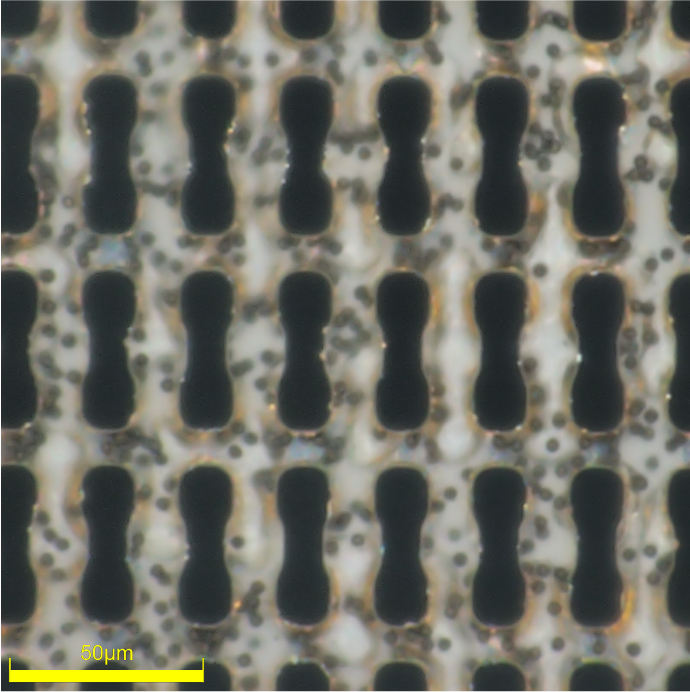}
    \caption{}
\end{subfigure}
\hfill
\begin{subfigure}{0.3\textwidth}
    \centering
    \includegraphics[width=\textwidth]{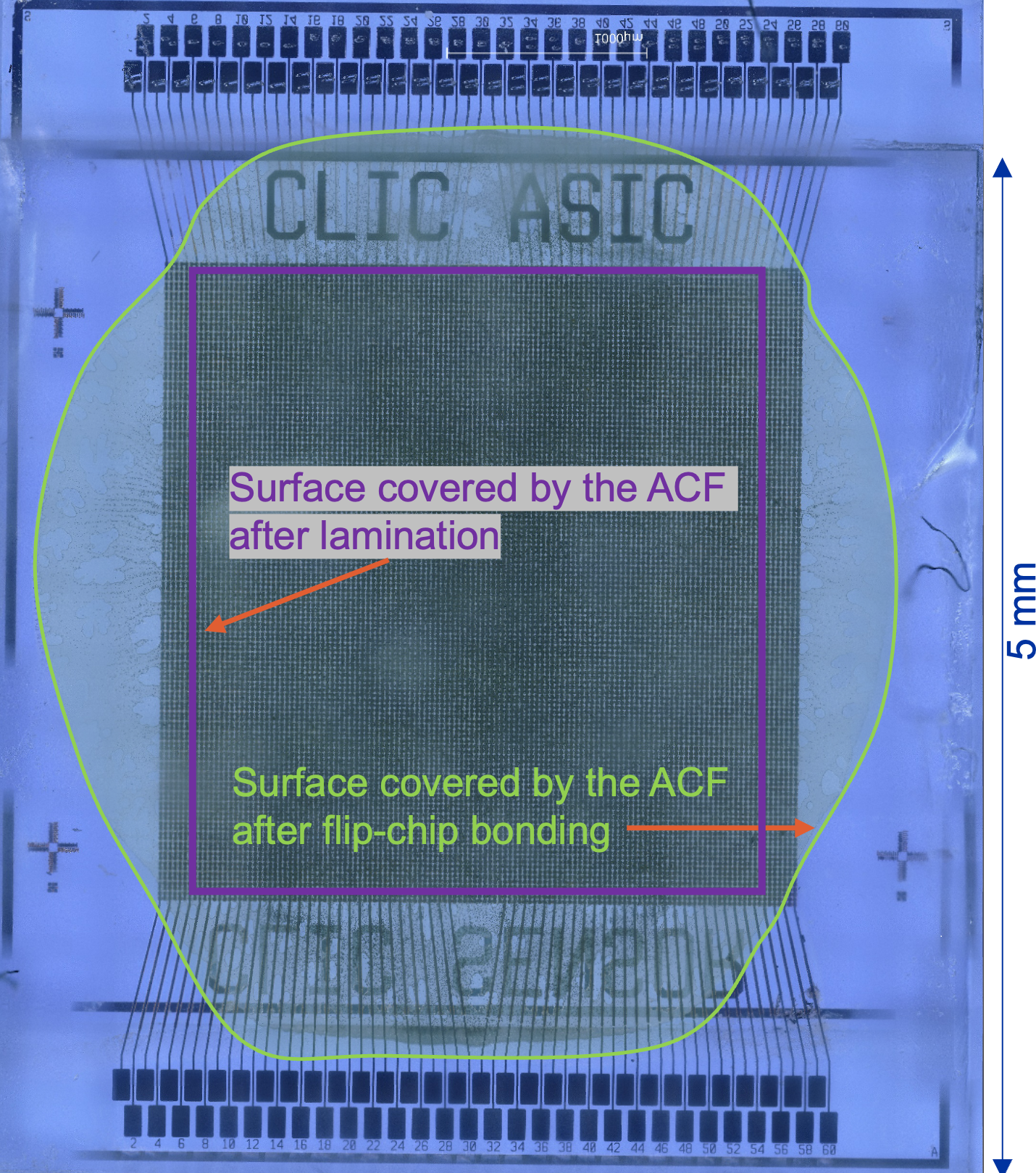}
    \caption{}
\end{subfigure}
\hfill
\caption{Optical-microscope characterisation pictures of the test-chip assembly  a) Bonding interface after separation of the chips (SEM), b) Test assembly with particles between the chains, view from the top, c) Test assembly with excess ACF outside the matrix.}
\label{fig:4}
\end{figure}

The flip-chip bonding of the two test chips was performed at 150~°C  during few seconds, under a force of 15~kgf, resulting in a mechanically stable assembly.
Electrical continuity tests could not be performed, since the probing lines on the test chips were damaged due to fabrication issues. Therefore, another method was employed to assess the connection yield: the chips were separated, and the bonding interface was inspected to evaluate the percentage of bumps on which at least one conductive particle had been smashed, indicating a connection after flip-chip bonding (Fig. \ref{fig:4}a).

Chip separation was performed using a pull tester, and a force of 11.4 N (corresponding to 7.2~kgf/cm² for this assembly) was needed to detach the ACF. Based on SEM characterisations, a connectivity yield of $\approx90~\%$ was estimated, which illustrates the potential of the ACF interconnect method for such fine-pitch applications.

Investigations were conducted to understand how the connection yield could be improved. Inspection revealed that, after lamination, conductive particles were uniformly distributed across all the surface of the chip as expected (Fig. \ref{fig:3}c). However, post-bonding analysis indicated significant particle misalignment, with some regions exhibiting voids and others clusters of particles (Fig. \ref{fig:4}b). This behaviour is attributed to the large thickness of the ACF film (18~µm), which is not suitable for fine-pitch bumps with a height of 4.5~µm. During the flip-chip bonding step, the excess ACF volume was displaced laterally toward the edges of the assembly, resulting in particle displacement and misalignment (Fig. \ref{fig:4}c).

In conclusion, this test represents our first attempt at connecting 25~µm-pitch test structures using ENIG bumping and ACF bonding, and the estimated connection yield is $\approx90~\%$. The results highlight the importance of matching the ACF thickness to the bump geometry and suggest that thinner ACF films, specifically designed for fine-pitch assemblies, can enable higher interconnection yields in future iterations.

\section{Development of a Program to Optimise ENIG Bump Height}

To improve the connection yield, an optimisation of the bump height with respect to the ACF thickness is required, taking into account the pitch and pad dimensions of the chips.
As shown above, a too large ACF thickness compared to the combined height of the bumps on both chips to be hybridised leads to the excess ACF volume being squeezed out of the pixel-matrix area during the flip-chip bonding process.
On the other hand, since the ENIG bump growth is isotropic, excessively large bumps reduce the inter-bump volume, thereby limiting the space available for the ACF between adjacent bumps. Consequently, the bump height must be carefully adjusted to the ACF thickness, in order to achieve optimal mechanical bonding and high electrical connection yield.

\begin{figure}[htbp]
\centering
\begin{subfigure}{0.35\textwidth}
    \centering
    \includegraphics[width=\textwidth]{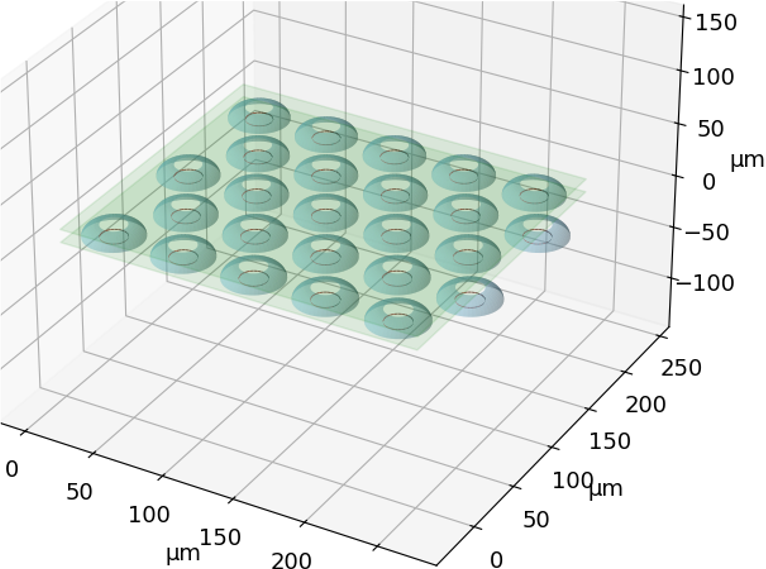}
    \caption{}
\end{subfigure}
\begin{subfigure}{0.45\textwidth}
    \centering
    \includegraphics[width=\textwidth]{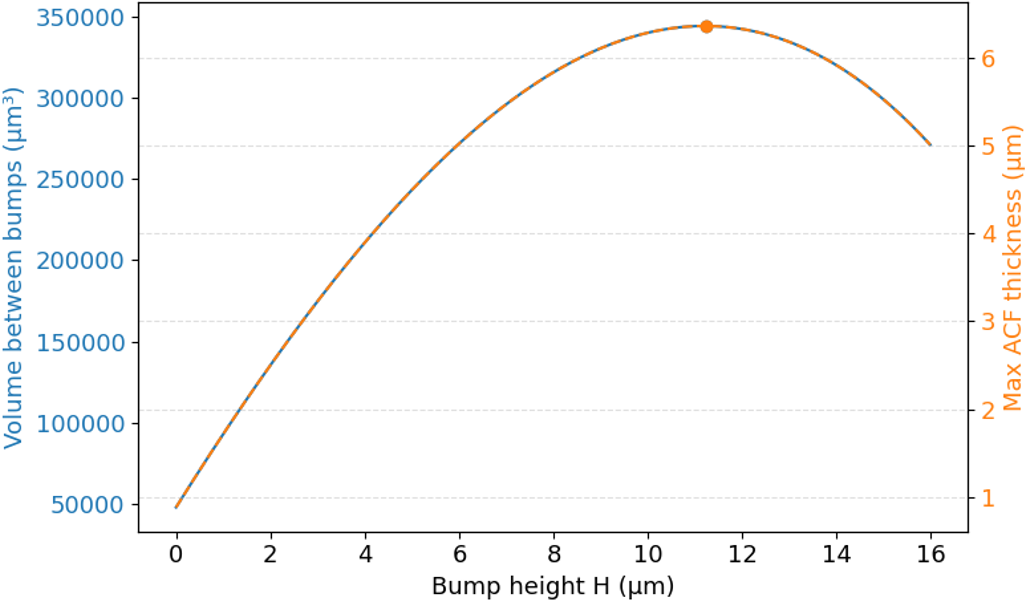}
    \caption{}
\end{subfigure}
\hfill
\caption{Simulation results for a 5 by 5 triangular matrix, with 18~µm passivation opening diameter and 50~µm pitch: a) 3D view of ENIG bumps, b) evolution of the inter-bump volume and the maximum compatible ACF thickness as a function of bump height.}
\label{fig:5}
\end{figure}

To address this, a dedicated simulation tool was developed to model the isotropic growth of ENIG bumps using analytical geometric equations and determine the optimum bump height based on chip parameters such as pad opening diameter, pitch, pattern of the pitch (triangular or square). Each ENIG bump is represented as a truncated dome with a flat top matching the pad diameter and curved sides representing isotropic growth. The simulator arranges the pads in either a square or triangular matrix according to user-defined rows, columns, and pitch. This tool was implemented in Python, using NumPy for numerical computation and Matplotlib for data visualization (2D and 3D plots). 
The program provides 3D visualisations of the simulated matrices (Fig. \ref{fig:5}a), as well as plots showing the evolution of the free volume between bumps and the maximum compatible ACF thickness as a function of bump height (Fig. \ref{fig:5}b).
This tool will provide guidance for adapting the ENIG bump height to the ACF thickness, thereby improving future hybridisation processes.

\section{Conclusion and Outlook}

This study presents results of ENIG–ACF flip-chip hybridisation for fine-pitch (55~µm and 25~µm) interconnections. The hybridisation of functional Timepix3 and Ti-LGAD sensors confirmed the process feasibility on functional components, achieving connection yields of 95\% with prospect for improvement. The hybridisation of 25~µm pitch interconnect test structures further extended the approach towards higher interconnect density, with a $\approx90~\%$ connectivity yield estimated.
The results highlight the importance of matching the ACF thickness with the bump height for achieving optimal electrical and mechanical bonding. Excessive ACF thickness was shown to induce particle displacement, reducing the effective connection yield. To mitigate this, a dedicated ENIG bump-height optimisation tool was developed, enabling predictive adjustment of bump geometry based on chip characteristics and ACF parameters.
Future work will focus on further optimising these processes to enhance connection yield, particularly for chips with pitches down to 25~µm. In addition, the impact of thermal cycling and irradiation on interconnect reliability will be investigated, with the goal of adapting this hybridisation method to the specific requirements of next-generation particle-physics detectors.

\acknowledgments
The developments presented in this contribution are performed within the DRD3 collaboration on Solid-State Detectors and in collaboration with the CERN EP R\&D programme on technologies for future experiments. This project has received funding from the European Union’s Horizon 2020 Research and Innovation programme under GA no 101004761.

\end{document}